# PROCYON-A AND ηBOOTIS:
# OBSERVATIONAL FREQUENCIES ANALYZED BY THE LOCAL-WAVE FORMALISM

**P.A.P. Nghiem** [(1)], **J. Ballot** [(2)], **R.A. García** [(1)], **P. Lambert** [(1)], **and S. Turck-Chièze** [(1)]

[(1)] *Service d'Astrophysique, DAPNIA/DSM/CEA CE Saclay*
*AIM – Unité mixte de recherche CEA-CNRS-Université Paris VII, UMR n°7158*
*91191 Gif sur Yvette Cedex, France*
[(2)] *Max Planck Institut für Astrophysik, Karl-Schwarzschild-Str. 1, Postfach 1317, 85741 GARCHING, Germany*
e-mail: papnghiem@cea.fr

**ABSTRACT**

In the present analysis of Procyon-A and ηBootis, we use the local-wave formalism which, despite its lack of precision inherent to any semi-analytical method, uses directly the model profile without any modification when calculating the acoustic mode eigenfrequencies. These two solar-like stars present steep variations toward the center due to the convective core stratification, and toward the surface due to the very thin convective zone. Based on different boundary conditions, the frequencies obtained with this formalism are different from that of the classical numerical calculation. We point out that (1) the frequencies calculated with the local-wave formalism seem to agree better with observational ones. All the frequencies detected with a good confident level including those classified as 'noise' find an identification, (2) some frequencies can be clearly identified here as indications of the core limit.

## 1. INTRODUCTION

After the important successes obtained by helioseismology to study the solar interiors, several observation campaigns have aimed to gather seismic informations on sun-like stars. Relatively consequent sets of p-mode frequencies are now available for the two solar-like stars Procyon-A and ηBootis, obtained from ground-based telescopes. Numerous theoretical interpretations of these observations have been done, but always following the same kind of numerical calculations. In spite of their very high precision, these calculations use boundary conditions different from those given by stellar models which lead to smooth out some physical variable profiles at the center and at the surface.

For ηBootis, ground-observed frequencies from a single site have been studied during the last decade: Kjeldsen et al. (1995); Christensen-Dalsgaard, Bedding & Kjeldsen (1995); Guenther & Demarque (1996); Di Mauro et al. (2003); Kjeldsen et al. (2003); Di Mauro et al. (2004); Guenther (2004); Carrier, Eggenberger & Bouchy (2005); Thévenin et al. (2005). And some first data obtained from the satellite MOST begin also to be studied: Guenther et al. (2005); Straka et al. (2006).

Concerning Procyon-A, essentially ground-based observations have been interpreted until now: Martíc et al. (1999); Barban et al. (1999); Chaboyer, Demarque & Guenther (1999); Provost et al. (2002); Martíc et al. (2004); Provost, Martíc & Berthomieu (2004); Eggenberger et al. (2004); Kervella et al. (2004); Claudi et al. (2005); Robinson et al. (2005); Eggenberger, Carrier & Bouchy (2005).

Those numerous works show the great interest of these frequency measurements despite their imprecision. The different stellar models obtained seem to converge toward similar parameters. But there are still debates on the mode identification, the coherence between calculated, ground-based, space-based frequencies, the possible coupling between p and g modes, etc. One can notice that all these studies use the same kind of numerical codes (adiabatic or non-adiabatic) to calculate the eigenfrequencies, with especially the same boundary conditions than those used for the Sun, despite typical disagreements with observations. Knowing that other boundary conditions should lead to noticeably different frequencies, we propose to confront the observed frequencies of the two stars with those obtained with the local-wave formalism, for which the main differences with usual calculations concern the boundary conditions.

In this paper, we first briefly recall some main principles of the local-wave formalism, its specificities, and its results for the Sun in section 2. Then we will compare its results to observations for the two stars Procyon-A and ηBootis in sections 3 and 4. Finally, some provisional conclusions will be drawn in section 5.

The stellar models used in this paper are obtained with the CESAM code (Morel, 1997). We do not search to build specific models to fit with seismic observations. We follow the above quoted authors, and simply search to obtain one model for each case, whose parameters are approximately within those given by them.

## 2. THE LOCAL-WAVE FORMALISM

The local-wave formalism is for the moment a semi-analytical approach which is discussed in details in Nghiem (2003a, 2006). This approach supposes that stellar oscillations are the manifestation of pure acoustic waves trapped inside a cavity. These waves can thus only propagate in a locally homogeneous environment, i.e. where the pressure scale-height is enough large compared to the wavelength. At the surface where pressure and density decrease rapidly, this condition will no longer fullfiled at a precise point, which determines the external turning point, different for each frequency. The phase there should be so that there is a displacement antinode. Toward the interior, the internal turning point is determined by the location where the wave number is cancelled, and the phase is calculated by a proper fitting with the Airy functions. The radial modes have a special treatment. As near the center, the spherical wave amplitude increases rapidly to infinity, this increase can be regarded as a decrease of environment pressure and everything can be treated exactly like for the external conditions.

This approach is a strict application of the JWKB approximation. It allows to treat stellar oscillations in a classical manner like for musical instruments, with a small number of equations. It has the advantage to use boundary conditions which avoids extra assumptions on the stellar structure. Unlike calculation methods that suppose either an abrupt limit or an isothermal atmosphere at the surface, and regular conditions toward the center, the local-wave concept uses the structure like it is, without any modification. The drawback is that g modes, along with coupling between p and g modes are not treated. Moreover, inherent to a semi-analytical method, the approximations used lead to larger uncertainties, ±3 µHz, than for numerical methods, << ±1 µHz. Nevertheless those uncertainties are the same for the whole frequency range, and the effect of boundary conditions amounts to tens of µHz.

To illustrate those specificities, let us consider the well known solar case. Frequencies for the modes $\ell = 0$ to 10 obtained by the present approach are represented in Fig. 1., along with numerical results (ADIPLS code, Christensen Dalsgaard, 1997 ), for the Saclay-seismic solar model (Turck-Chièze et al. 2001; Couvidat et al. 2003), compared to observed frequencies (Rhodes et al. 1997; Bertello et al. 2000; García et al. 2001). The differences between these three frequency sets are independent of the degree $\ell$. This shows that they come from the very near surface region, where all modes of any $\ell$ have the same trajectory. Finally, take note that, at a precision of ±3 µHz, frequencies obtained by the present method can be shifted to coincide with numerical ones when an isothermal atmosphere is adjusted to the external end of the standard solar model (Nghiem 2003b), and to coincide with observed ones when a precise magnetic profile is added to the solar-model surface (Nghiem et al. 2006).

## 3. PROCYON-A

The Procyon-A model used here has the following characteristics: mass $M = 1.497\ M_\odot$, radius $R = 2.045\ R_\odot$, luminosity $L = 6.879\ L_\odot$, Age = 1.77 Gy, no atomic diffusion, no core overshooting, Eddington type atmosphere, OPAL2001 equations of state, OPA-HOUDEK9 opacities, standard treatment of convection with the mixing-length parameter $\alpha = 1.4048$, initial mass fractions $X_0 = 0.71453$, $Y_0 = 0.26975$. This model describes a star in its late main-sequence stage, with a convective core that extends up to the relative radius $r/R = 0.057$, and a thin convective envelope which lies down to only $r/R = 0.95$. Its sound-speed profile is given in Fig. 2. Toward the center, an abrupt variation can be noted right at the core limit, due to the rapid change of the chemical composition. Toward the surface, steep variations occur around $r/R \sim 0.99$, induced by the thinness of the convective zone.

These two pronounced transitions should strongly affect the acoustic frequencies. Furthermore, it should be easy to distinguish between the two effects, because every surface effect is $\ell$-independent, while an abrupt change toward the center will affect only some specific modes that have their internal turning point around this location (Nghiem et al. 2004). That can be seen on the large separations presented in Fig. 3. The present calculation shows indeed wide undulations of a few µHz, independent of $\ell$, for the frequencies between 300 and 600 µHz, which have their external turning points at r / R = 0.980 to 0.997, whatever their degree; and a 3 µHz single peak for the mode $\ell = 1$, ν = 828 µHz which has its internal turning point at r / R = 0.047.

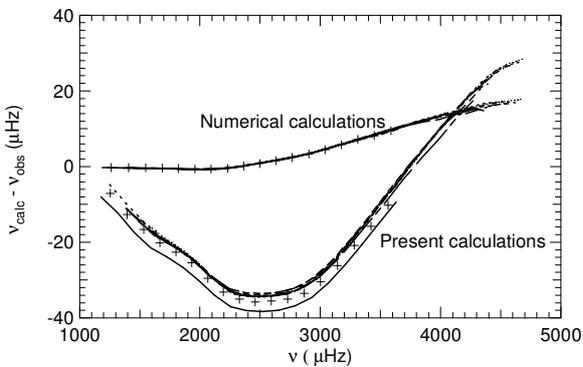

Fig. 1. Frequency differences for the Sun, calculations minus observations, for $\ell = 0$ to 10 modes. Radial modes are represented by crosses, $\ell = 1$ modes are joined by a continuous line, higher-$\ell$ modes by progressively discontinuous lines.

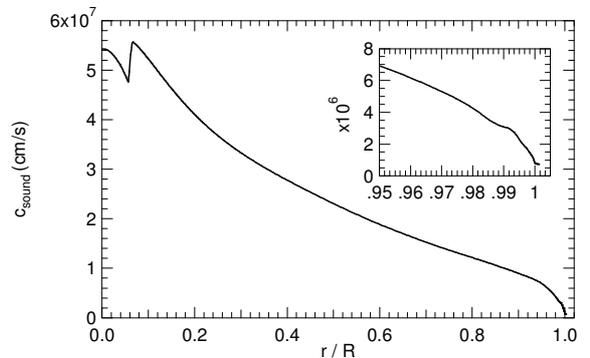

Fig. 2. Procyon-A sound-speed profile and zoom at the surface.

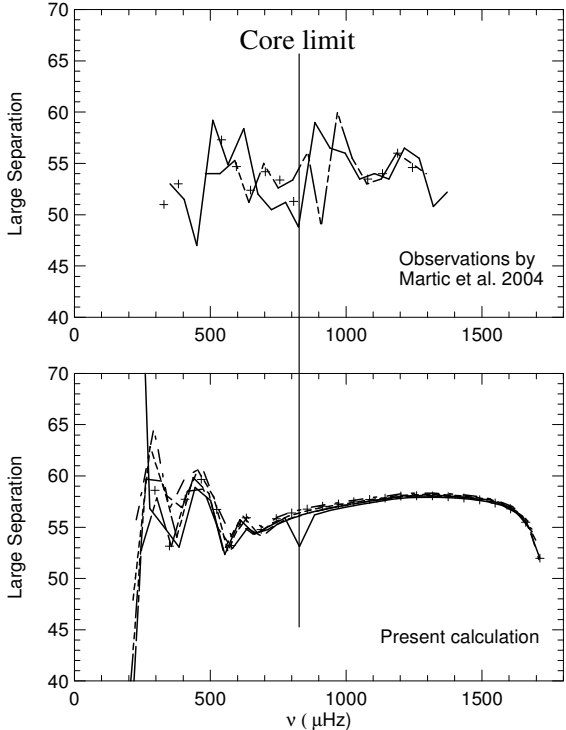

Fig. 3. *Upper panel: large separations for the modes $\ell = 0, 1, 2$ obtained with the 55 observed frequencies by Martíc et al. 2004.*
*Lower panel: large separations for the modes $\ell = 0$ to 5, obtained with the present calculation. Symbols as defined in Fig.1.*

The large separations calculated with the 55 observed frequencies by Martíc et al. (2004) are plotted on the same Fig. 3. Referring to numerical calculations, the peaks of $\ell = 1$ modes in the range 450 to 826 µHz are attributed by many authors to possible couplings with g modes. But numerical calculations predict for this degree a coupling rather at 325 µHz, inducing a much more important shift of -30 µHz. These observational undulations at low frequency seem more compatible with the $\ell$-independent undulations, of a few µHz amplitude, obtained with the local-wave formalism. Moreover, the $\ell = 1$ peak at 826 µHz corresponds well to that indicating the core limit, predicted by the latter. Martíc et al. (2004) observed also a cut-off frequency of ~1700 µHz, which is more compatible with the estimation following the local-wave concept (1725 µHz), than that using an isothermal atmosphere (1583 µHz).

Unfortunately these observational data are not enough precise to draw firm conclusions. One can only speak about a more or less compatibility with theoretical predictions. Furthermore, the day/night effect inherent to single-site observation introduces an alias that 'allows' to arbitrary modify the frequencies by ± 11.57µHz, which contributes to recall us to be more cautious. As the paper in question mentions that such corrections have been applied but does not indicate which frequencies have been corrected, we cannot analyse them further.

Let us switch now to observational frequencies obtained by Eggenberger et al. (2005). There are 23 frequencies, plus 5 frequencies that are classified as 'noises', although their signal/noise is equivalent to the others, because they do not correspond to numerical predictions. In this study the original detected frequencies before the day/night corrections are clearly indicated, so we can search to apply these corrections. As these operations do not have other justifications than trying to make observation-theory coincide, we cannot totally trust in them, and will not attempt further to identify the frequencies found with any degree $\ell$. From now on, instead of showing large separation diagrams, we will show only echelle diagrams.

Frequencies obtained by the present calculation, for $\ell = 0, 1, 2, 3$, and observed frequencies for Procyon-A are plotted in the echelle diagram of Fig. 4. As indicated in Table 1, certain observed frequencies are shifted by the day/night alias. We have also included the 5 frequencies originally classified as 'noise' by their authors (marked by an asterisk). We can see that there is a global agreement between calculated and observed frequencies within ±5 µHz. As compared, Eggenberger et al. (2005), using different day/night shifts, obtained the same kind of agreement with numerical results, but after eliminating 5 frequencies and ignoring an $\ell$-independent disagreement of a few µHz.

Differences between the present calculation and traditional numerical calculations come from the different boundary conditions used. Surface conditions induce in the echelle diagram a different global trend common to every degree, i.e. the undulations at low frequencies, and a much more inclined pattern at high frequency. This reflects directly the physical conditions at the star-model surface. As observations appear to conform to this scheme, it seems therefore that unlike for the Sun, the surface modelisation by classic stellar evolution code is realistic for Procyon-A.

The effects of abrupt changes at the core limit can be seen by following the $\ell = 1$ line, i.e. the open squares in Fig. 4, from low to high frequencies. Only for this degree, there is a step of 2 µHz at 828 µHz, and precisely this frequency has been observed (see Table 1). Unfortunately the frequencies immediately lower and higher have not been observed. Like with the data of Martíc et al. (2004), this predicted core-limit effect seems therefore to be fully compatible with observations, but still remains to be precisely confirmed.

*Table 1. Procyon-A. The 23 + 5 observed frequencies of Eggenberger et al. (2005), shifted differently than the authors, by the day/night alias.*

| Frequency ( µHz) |
|---|
| 651.5 + 11.57 = 663.1 |
| 630.8 |
| 662.7 * |
| 683.5 + 11.57 = 695.1 |
| 720.6 |
| 797.9 |
| 799.7 |
| 791.8 + 11.57 = 803.4 |
| 828.5 |
| 835.4 – 11 .57 = 823.8 * |
| 856.2 |
| 859.8 |
| 911.4 |
| 929.2 – 11.57 = 917.6 |
| 929.2 + 11.57 = 940.8 |
| 1009.7 – 11.57 = 998.1 |
| 1027.1 |
| 1123.3 – 11.57 = 1111.7 |
| 1137.0 + 11.57 = 1148.6 |
| 1131.1 + 11.57 = 1142.7 |
| 1192.4 + 11.57 = 1204.0 |
| 1186.0 + 11.57 = 1197.6 |
| 1234.8 + 11.57 = 1246.4 |
| 1251.8 + 11.57 = 1263.4 |
| 1265.6 |
| 1337.2 * |
| 1439.0 * |
| 1559.5 – 11.57 = 1547.9 * |

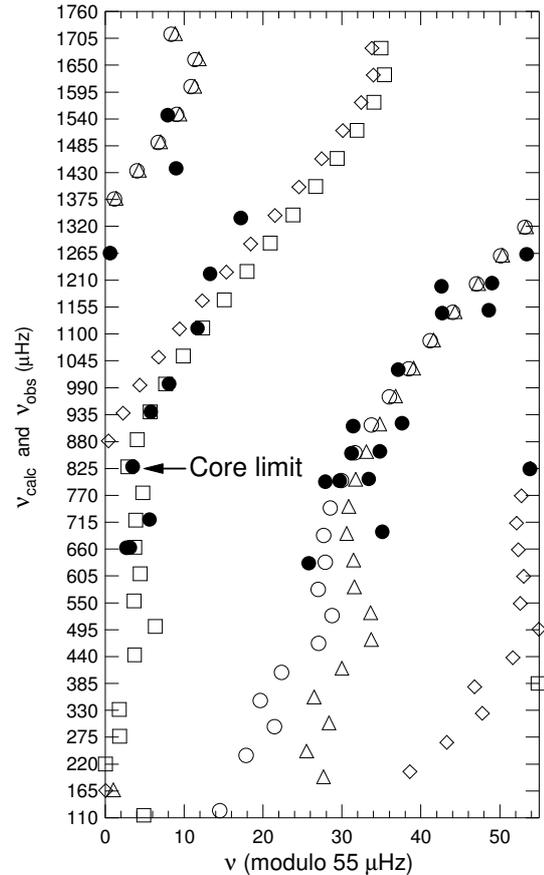

*Fig. 4. Ehelle diagram for Proyon-A. Open symbols are calculated frequencies for $\ell = 0$(circle), 1(square), 2(triangle), 3(diamond). Filled circles are the 23 + 5 observed frequencies of Eggenberger et al. (2005), shifted differently than the authors, as indicated in Table 1.*

Other effects of the convective core deserve to be noted. The abrupt change at the core limit induces strong differences on cavity lengths between modes that have their internal turning point inside or outside the core. We have already noted that the $\ell = 1$ modes have their internal limit outside the core at low frequency, and inside at high frequency. This fact leads to a step in the echelle diagram. This implies in addition that the high frequency part will have suddenly a larger cavity, thus shift to lower frequencies, very close to the $\ell = 3$ line, of which all the modes have their turning point outside the core. Concerning now the modes $\ell = 0$ and 2, the first ones have all their turning points inside the core, and the second ones outside. Compared to the case without convective core, the radial modes will have larger cavities, thus shift to lower frequencies, so that its curve in the echelle diagram appear on the left of the $\ell = 2$ one, unlike what happens for the Sun. This effect is less pronounced for high frequencies which have anyway their turning points very near the center. That is why the frequency distance between the curves $\ell = 0$ and 2 (the small difference) will decrease to merge at high frequency. This special behaviour has been pointed out by Martíc et al. (2004) when looking at their observational frequencies above 700 µHz, and can be clearly seen in Fig. 6 and 8 of their paper. This progressive merging is not predicted by numerical frequency calculations where regular conditions are applied near the center, for which the curves $\ell = 0$ and 2 appear nearly parallel in the echelle diagram.

**4. ηBOOTIS**

Adopting averaged values for the parameters resulting from the works performed by the authors cited in the introduction part, and more especially Thévenin et al. (2005), we study the two following models of ηBootis:

- Without core overshooting. Mass $M = 1.7$ $M_\odot$, radius $R = 2.71$ $R_\odot$, luminosity $L = 8.83$ $L_\odot$, Age = 2.485 Gy, no atomic diffusion, Kurucz atmosphere, OPAL2001 equations of state, OPA-HOUDEK9 opacities, standard treatment of convection with the mixing-length parameter $\alpha = 1.8$, initial mass fractions $X_0 = 0.71$, $Y_0 = 0.25$. This model describes a star that has just left the main sequence. The central hydrogen burning is just finished, so the core has turned from convective into radiative, but still keeps its former stratification with a strongly marked edge at $r/R = 0.038$. The convective envelope begins its downward expansion, and reaches $r/R = 0.83$.
- With core overshooting, $\alpha_{ov} = 0.15$, and atomic diffusion. Mass $M = 1.7$ $M_\odot$, radius $R = 2.71$ $R_\odot$, luminosity $L = 8.85$ $L_\odot$, Age = 2.982 Gy. Any other physics ingredients staying the same than in the previous model, the core edge is now larger, at $r/R = 0.059$.

For these two models, the sound-speed profile and the large separations for modes $\ell = 0$ to 5 are given in Fig. *5* and *6*. Like with Procyon-A, steep variations occur toward the surface, at $r/R = 0.985$, which induce a wide undulation at low frequency in the large-separation curves for every degree. Toward the center, as expected with the overshooting character, the abrupt variation of the sound speed is more strongly marked and expands less far from the center for the case without overshooting than for the case with overshooting. This leads to important frequency differences between the two cases. In the first case, only the radial oscillation modes have their internal turning points inside the core, and very near the center, while every other mode does not penetrate into the core. In the second case, radial modes have their turning points inside the core but less near the center, and modes of degree $\ell = 1$ with frequencies higher than 812 µHz have their turning points inside the core, while the others have them outside. That is why only the large separation curve of $\ell = 1$ for the model with overshooting presents a special peak at 812 µHz, and this is a large peak of 6 µHz. Like with Procyon-A, only more pronounced, one notes in the echelle diagram an important step separating the $\ell = 1$ curve into two shifted-away parts. In the same diagram, the radial modes will also be shifted strongly to lower frequency, because their cavities are larger, and this much more for the case without overshooting.

Those behaviors can indeed be seen in the echelle diagram of Fig. 7 and 8. The $\ell = 0$ curve, unlike for the Sun where it is located on the right of the $\ell = 2$ curve, has been shifted completely on the left on the diagram when there is no overshooting, and less shifted when there is overshooting. Only for this last case, a step at 812 µHz can be seen clearly when following the $\ell = 0$ curve (open squares) from low to high frequency. On these diagrams, we have also reported the observed frequencies composed of: 22 + 5 'noises' of Kjeldsen et al. (2003), 22 + 1 'noise' of Carrier et al. (2005), and 15 from the MOST satellite (Guenther et al. 2005). The ground-based frequencies have been shifted by the day/night alias of ± 11.57μHz as indicated in Table 2 and 3, in order to make them as close as possible to calculated frequencies. The asterisks mark the frequencies originally classified as 'noise' by the authors because no corresponding numerical frequencies were found.

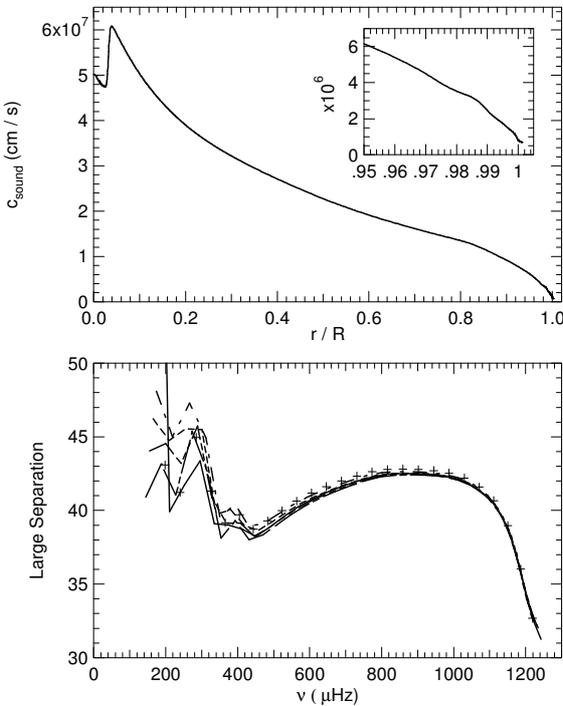

*Fig. 5. ηBootis without core overshhooting. Upper panel: sound speed profile and zoom at the surface. Lower panel: large separations for the modes $\ell = 0$ to 5. Symbols as defined in Fig.1.*

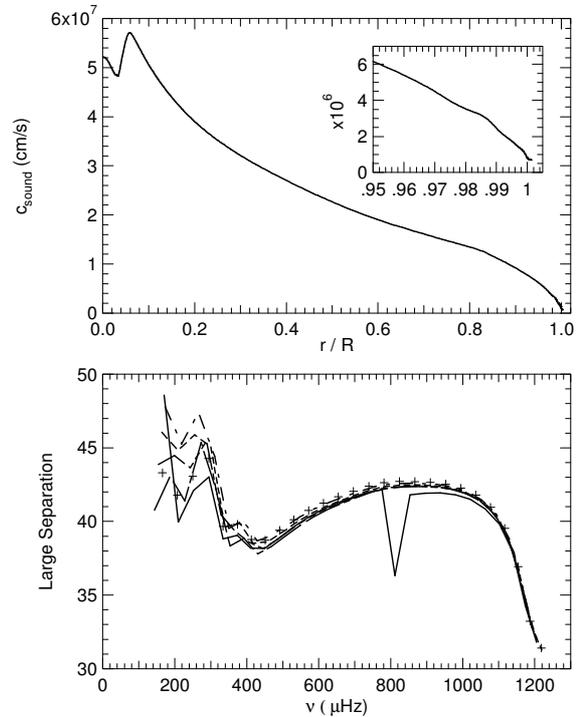

*Fig. 6. ηBootis with core overshhooting. Upper panel: sound speed profile and zoom at the surface. Lower panel: large separations for the modes $\ell = 0$ to 5. Symbols as defined in Fig.1.*

The calculation/observation disagreements are at maximum 4 μHz in both Fig. 7 and 8, where all the three frequency sets are considered, including the so-called 'noise' frequencies. But the agreement appears well better when there is overshooting: the $\ell = 0$ curve is less shifted to low frequencies, and comes very close to the Most observations. The curious bifurcation of the latter at 211 μHz can even be explained by the crossing of the $\ell = 0$ and 1 curves. And the important step of 3.5 μHz in the middle of the $\ell = 0$ curve seems also to be well confirmed, indicating that the core limit can be very easy to detect.

Does this mean that, unlike with the Sun but like with Procyon-A, the stellar model does not present important differences at the surface so that calculated and observed frequencies can be directly comparable? The situation was not so clear when numerical calculations were considered with an isothermal-atmosphere fitting at the surface. Generally (see references in the introduction part), a reasonable agreement was obtained with only one of the ground-based observations, after having discarding the so-called 'noise' frequencies, and removing a global disagreement of a few μHz. A more recent study of Straka et al. (2006) finds that an appropriate introduction of turbulence together with non-adiabaticity at the surface can reduce partly these discrepancies. But the authors notice also that the structure change reproducing the correct shift of the p-mode frequencies is not unique.

Table 2. η Bootis without core overshooting. The 22 + 5 observed frequencies of Kjeldsen et al. (2003), and the 22 + 1 frequencies of Carrier et al. (2005), shifted differently than the authors, by the day/night alias.

| Frequency ( μHz) ||
|---|---|
| Kjledsen et al. (2003) | Carrier et al. (2005) |
| 611 | 512.2 + 11.57 = 523.8 |
| 639.6 + 11.57 = 651.2 | 544.6 – 11.57 = 533.0 |
| 679.2 + 11.57 = 690.8 | 550.3 |
| 732.6 | 589.9 |
| 813.1 | 622.2 – 11.57 = 610.6 |
| 853.6 | 614.1 – 11.57 = 602.53 |
| 905.8 | 653.8 |
| 975.7 + 11.57 = 987.3 | 669.9 |
| 641.0 – 11.57 = 629.4 | 691.3 |
| 670.1 | 724.5 – 11.57 = 712.9 |
| 711.8 | 728.3* |
| 719.3 – 11.57 = 737.7 | 729.5 |
| 765.0 – 11.57 = 753.4 | 748.5 – 11.57 = 736.9 |
| 793.1 – 11.57 = 781.5 | 777.2 |
| 955.6 – 11.57 = 944.0 | 781.0 – 11.57 = 769.4 |
| 1034.3 | 775.8 |
| 608.1 | 805.1 + 11.57 = 816.7 |
| 716.8 + 11.57 = 728.4 | 809.2 – 11.57 = 797.6 |
| 810.5 | 834.5 – 11.57 = 822.9 |
| 849.9 + 11.57 = 861.5 | 888.7 + 11.57 = 900.3 |
| 960.1 – 11.57 = 948.5 | 891.6 + 11.57 = 903.2 |
| 962.0 – 11.57 = 950.4 | 947.6 |
| 657.1 – 11.57 = 645.5* | 960.3 – 11.57 = 948.7 |
| 665.8 – 11.57 = 654.2* | |
| 806.7 – 11.57 = 795.1* | |
| 815.9* | |
| 1070.4* | |

Table 3. η Bootis with core overshooting. Same frequencies as Table 2, but certain of them shifted differently.

| Frequency ( μHz) ||
|---|---|
| Kjledsen et al. (2003) | Carrier et al. (2005) |
| 611 | 512.2 + 11.57 = 523.8 |
| 639.6 + 11.57 = 651.2 | 544.6 – 11.57 = 533.0 |
| 679.2 + 11.57 = 690.8 | 550.3 + 11.57 = 561.9 |
| 732.6 | 589.9 |
| 813.1 | 622.2 – 11.57 = 610.6 |
| 853.6 | 614.1 |
| 905.8 – 11.57 = 894.2 | 653.8 |
| 975.7 – 11.57 = 964.1 | 669.9 |
| 641.0 – 11.57 = 629.4 | 691.3 |
| 670.1 | 724.5 – 11.57 = 712.9 |
| 711.8 | 728.3 + 11.57 = 739.9* |
| 719.3 – 11.57 = 737.7 | 729.5 + 11.57 = 741.1 |
| 765.0 – 11.57 = 753.4 | 748.5 – 11.57 = 736.9 |
| 793.1 – 11.57 = 781.5 | 777.2 |
| 955.6 + 11.57 = 967.2 | 781.0 |
| 1034.3 – 11.57 = 1022.7 | 775.8 |
| 608.1 | 805.1 + 11.57 = 816.7 |
| 716.8 + 11.57 = 728.4 | 809.2 – 11.57 = 797.6 |
| 810.5 | 834.5 – 11.57 = 822.9 |
| 849.9 – 11.57 = 838.3 | 888.7 – 11.57 = 877.1 |
| 960.1 – 11.57 = 948.5 | 891.6 – 11.57 = 880.0 |
| 962.0 – 11.57 = 950.4 | 947.6 – 11.57 = 936.0 |
| 657.1 – 11.57 = 645.5* | 960.3 – 11.57 = 948.7 |
| 665.8 – 11.57 = 654.2* | |
| 806.7 – 11.57 = 795.1* | |
| 815.9* | |
| 1070.4 – 11.57 = 1058.8* | |

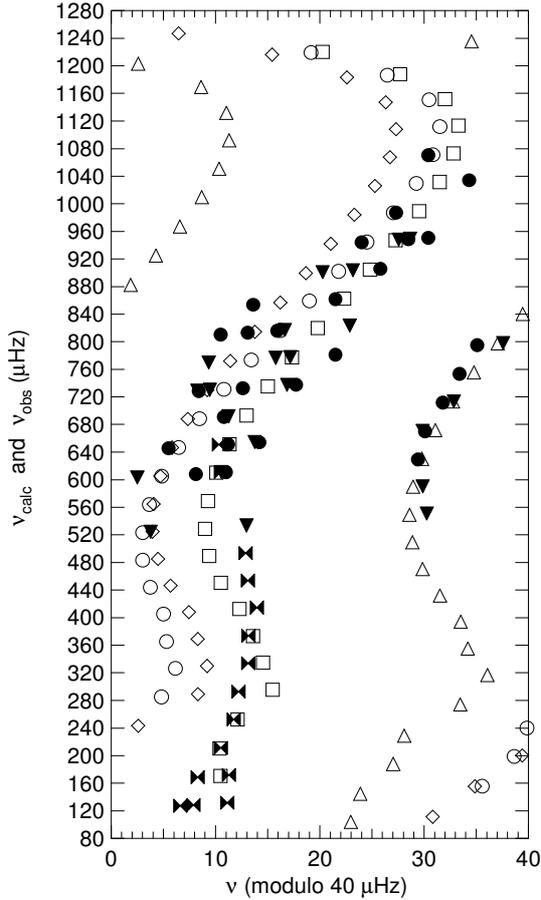 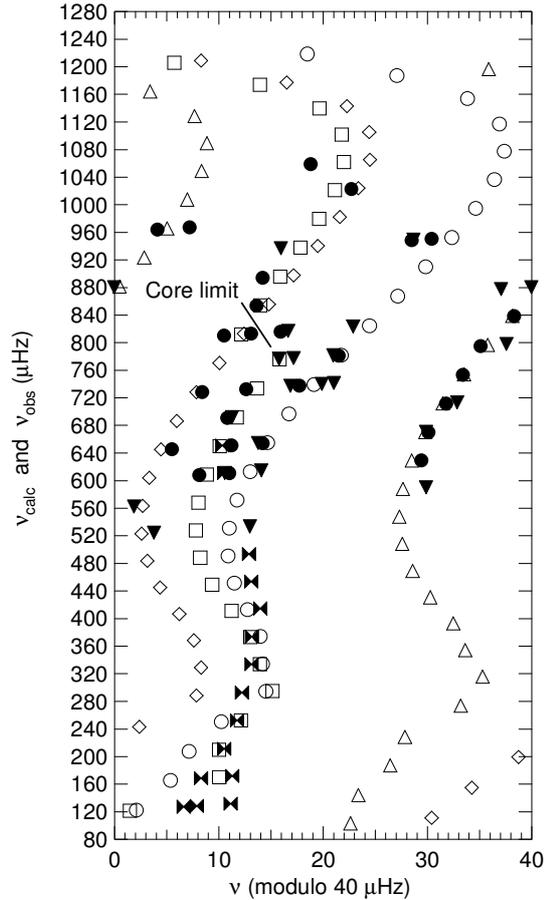

*Fig. 7. Echelle diagram for ηBootis without core overshooting.*   *Fig. 8. Echelle diagram for ηBootis with core overshooting.*

*Open symbols are calculated frequencies for ℓ = 0(circle), 1(square), 2(triangle), 3(diamond). Filled circles are the 22 + 5 observed frequencies of Kjeldsen et al. (2003), filled triangles are the 22 + 1 frequencies of Carrier et al. (2005), filled bow ties are the 15 Most frequencies (Guenther et al. 2005). Ground-based frequencies are shifted as indicated in Table 2 and 3.*

## 5. DISCUSSIONS AND CONCLUSIONS

Using different boundary conditions in the p-mode calculation leads to noticeably different oscillation frequencies. External condition effects are independent of the degree $\ell$ (for the low $\ell$ considered here), while internal conditions influence frequencies of each degree $\ell$ differently. The present work is based on the local-wave formalism employing boundary conditions that are different from traditional calculations. The advantage is that the stellar structure is not modified when the frequency calculation is performed.

We note a better agreement between the predicted frequencies and available measurements for the two studied stars. All measurements have been taken into account, including the frequencies classified as 'noise' but having the same signal/noise as others. No $\ell$-independent frequency shift is needed to make calculations coincide with observations. If this is confirmed, by thinking that for the Sun the frequency discrepancies can be explained by a surface magnetic field (Nghiem et al. 2006), we would conclude that the magnetic pressure at the surface of these two stars should not be a very preponderant component.

Otherwise, the coupling with g modes is not treated in the present approach. It seems therefore that it is not necessary to invoke this coupling to explain the available observed frequencies. Many authors have suspected such couplings when apparent discrepancies occured for non-radial modes, although they recognize that these discrepancies do not appear at the right frequencies nor induce the expected shift. According to the present work, these observations can be naturally explained by the two effects of the rapid change of the sound speed near the center: a strong shift of $\ell = 0$ modes toward lower frequencies, and a gap for the two successive frequencies of $\ell = 1$ modes which have their internal turning points on the two sides of this abrupt change. If this is confirmed, this would demonstrate the power of seismology to directly 'see' the deeply-burried stellar core, and by the way to bring up a material proof to modelisation theories of intermediate-mass stars.

These results seem very promising for the local-wave formalism. But great caution must be kept in mind. Single-site measurements allow too wide interpretation possiblities with the day/night alias. It is possible that some of the frequencies we have shifted by ± 11.57μHz, should not be shifted, or are shifted in the wrong direction. Space observations avoid this drawback but up to now very few are available, and they still concern a small range of frequencies. Networks of ground-based measurements and more complete space observations are necessary to definitely conclude on these points.

At the present stage, we would like to draw attention on the fact that different frequency calculations exist, arriving to different frequency sets. Interpreting observed frequencies following a theoretical scheme rather than another, could lead to different results. We would recommend to observers to publish also their raw frequencies before any theoretical interpretations.